\documentclass[a4paper,11pt]{article}
\pdfoutput=1 

\usepackage{jheppub} 

\usepackage[T1]{fontenc} 

\usepackage{amsfonts,amsmath,amssymb}
\usepackage{graphicx} 
\usepackage{booktabs}
\usepackage{hyperref}
\usepackage{mathtools}
\usepackage{float}
\usepackage[nottoc, notlot, notlof]{tocbibind}
\usepackage{physics}
\usepackage{bigints}
\usepackage[makeroom]{cancel}
\usepackage{braket}
\usepackage[T1]{fontenc}
\usepackage{color}
\usepackage{graphicx}
\usepackage{float}
\usepackage{chngpage}
\usepackage{ragged2e}

\allowdisplaybreaks

\preprint{TUM-HEP-1385/21}

\title{\boldmath Planar three-loop master integrals for
$2 \to 2$ processes with one external massive particle}

\author[a,b]{Dhimiter D. Canko}
\author[a,c]{and Nikolaos Syrrakos}
\affiliation[a]{Institute of Nuclear and Particle Physics, NCSR "Demokritos",\\ Agia Paraskevi 15310, Greece}
\affiliation[b]{Department of Physics, University of Athens,\\Zographou 15784, Greece}
\affiliation[c]{Physik-Department, Technische Universität München,\\ James-Franck-Str. 1, 85748 Garching, Germany}
\emailAdd{jimcanko@phys.uoa.gr}
\emailAdd{nikolaos.syrrakos@tum.de}

\abstract{We present analytic results for the two tennis-court integral families relevant to $2\to2$ scattering processes involving one massive external particle and massless propagators in terms of Goncharov polylogarithms of up to transcendental weight six. We also present analytic results for physical kinematics for the ladder-box family and the two tennis-court families in terms of real-valued polylogarithmic functions, making our solutions well-suited for phenomenological applications.}


\begin{document} 
\maketitle
\flushbottom

\section{Introduction}
Having a solid understanding of the physics of strong interactions is of vital importance for the analysis of experimental data coming from high-energy proton-proton colliders, such as the LHC \cite{Gehrmann:2021qex}. The increasing precision of these experiments translates to a high demand of equally precise theoretical predictions. Such calculations are performed in the framework of perturbative Quantum Field Theory (QFT), and in the case of strong interactions, perturbative Quantum Chromodynamics (QCD). Obtaining more precise predictions amounts to the computation of higher order corrections for QCD dominated processes, which involves, among other things, the computation of multiloop Feynman diagrams. Using the Feynman rules of QCD one can translate these diagrams to the so-called Feynman integrals (FI), which are then computed either numerically or analytically in the framework of dimensional regularisation in $d=4-2\epsilon$ space-time dimensions.  

It has been estimated that the scheduled High Luminosity upgrade of the LHC will require the calculation of many scattering processes at the Next-to-Next-to-Next-to-Leading-Order (N3LO) to reach per-cent level of precision \cite{Amoroso:2020lgh}. At that order of perturbative expansion the computation of three-loop FI becomes a necessity, a highly non-trivial task despite the major breakthroughs that have occurred in our understanding of their mathematical structure and the development of many methods for their analytical and numerical calculation \cite{Heinrich:2020ybq, Tancredi:2021oiq}. Currently, at N3LO all FI for $2\to2$ scattering involving massless propagators and on-shell external particles have been computed analytically \cite{Henn:2013tua, Henn:2020lye}, and the first phenomenological studies have appeared in the literature \cite{Caola:2020dfu, Caola:2021rqz, Bargiela:2021wuy,Caola:2021izf}. These impressive results have greatly benefited from our ability to compute FI analytically using the method of differential equations (DE) \cite{de1, de2, de3, de4}. The extension of these results to processes involving one massive external particle, such as a vector boson decaying to 3-jets or $gg\to H + jet$ in gluon fusion, is a natural next step.

It has been twenty years since the calculation of the two-loop FI for $2\to2$ processes involving massless propagators and one off-shell external particle \cite{Gehrmann:2000zt, Gehrmann:2001ck} which established the method of DE as a powerful technique for the analytic computation of multiloop FI. These studies developed an approach which in essence is still used today, i.e. using Integration-By-Parts (IBP) identities \cite{Chetyrkin:1981qh, Laporta:2000dsw} to identify a minimal set of FI, known as \textit{master integrals} (MI), deriving DE for these integrals by differentiating with respect to kinematic invariants and, if present, internal masses, and then using IBP identities again to recast the derivatives of the MI in terms of MI of equal or lower number of propagators.

A few years ago, a first step was taken towards the extension of these results at the three-loop level, with the calculation of the so-called planar ladder-box topology involving massless propagators and one off-shell leg \cite{DiVita:2014pza}. This calculation was made possible through the adoption of the by now established method of canonical DE \cite{Henn:2013pwa}. In \cite{Henn:2013pwa} an observation was made that one can choose a \textit{good} basis of MI, such that the DE that they satisfy have only logarithmic singularities and the dimensional regulator $\epsilon$ is fully factorised. The conjecture made in \cite{Henn:2013pwa} was that FI with constant leading singularities \cite{Arkani-Hamed:2010pyv} are good candidates for the construction of such a basis, which is known as a \textit{pure basis} of MI.

In \cite{DiVita:2014pza} a basis of 85 MI was obtained and a canonical DE was derived based on Magnus series expansions \cite{Argeri:2014qva}. In \cite{Canko:2020gqp} we re-calculated the three-loop ladder-box topology using a variant of the standard DE method, known as the Simplified Differential Equations (SDE) approach \cite{Papadopoulos:2014lla}. Using modern IBP tools such as \texttt{FIRE6} \cite{Smirnov:2019qkx} and \texttt{Kira2} \cite{Klappert:2020nbg}, we obtained a set of 83 MI and verified through the analytic results of \cite{DiVita:2014pza} that this is indeed the minimal set of integrals needed for the computation of all Feynman diagrams contained in the three-loop ladder-box family. 

In this paper we take a step further and compute fully analytically the two additional planar three-loop topologies, known as tennis-court topologies, in terms of Goncharov poloylogarithms (GPLs) \cite{Goncharov:1998kja, Duhr:2011zq, Duhr:2012fh, Duhr:2014woa} of up to transcendental weight six. We also provide expressions for all three planar integral families in all physical regions of the phase-space, thus allowing our results to be readily available for use in phenomenological studies. More specifically, in section \ref{sec:families} we introduce the integral families treated in this paper, the SDE parametrisation and set up basic notation, in section \ref{sec:cde} we construct pure bases for the two tennis-court families, derive canonical SDE for these bases and compute the relevant boundary terms, in section \ref{sec:results} we discuss the analytic continuation of our results into physical regions and present an analysis concerning the GPLs involved in our solutions and their numerical evaluation, as well as report on the various checks that we have performed to validate our solutions. Finally, we give our closing remarks in section \ref{sec:conclusion}. In appendix \ref{sec:appA} we comment on the adjacency conditions that were recently discovered in \cite{Dixon:2020bbt, Chicherin:2020umh} regarding the symbol of two-loop and three-loop four-point master integrals with one off-shell leg. 

\section{General set up}
\label{sec:families}
\subsection{Integral families}
\begin{figure} [h!]
\centering
\includegraphics[width=9cm]{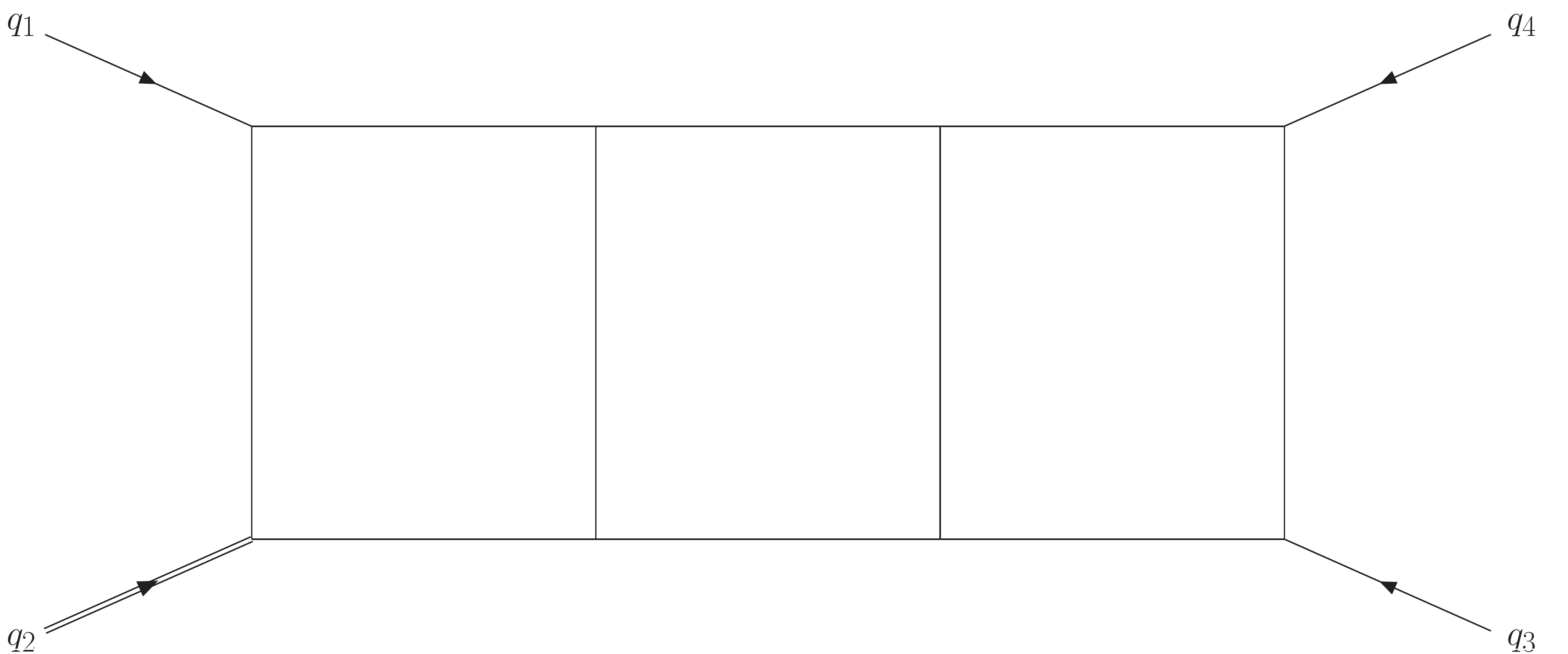}\\
\includegraphics[width=5.4cm]{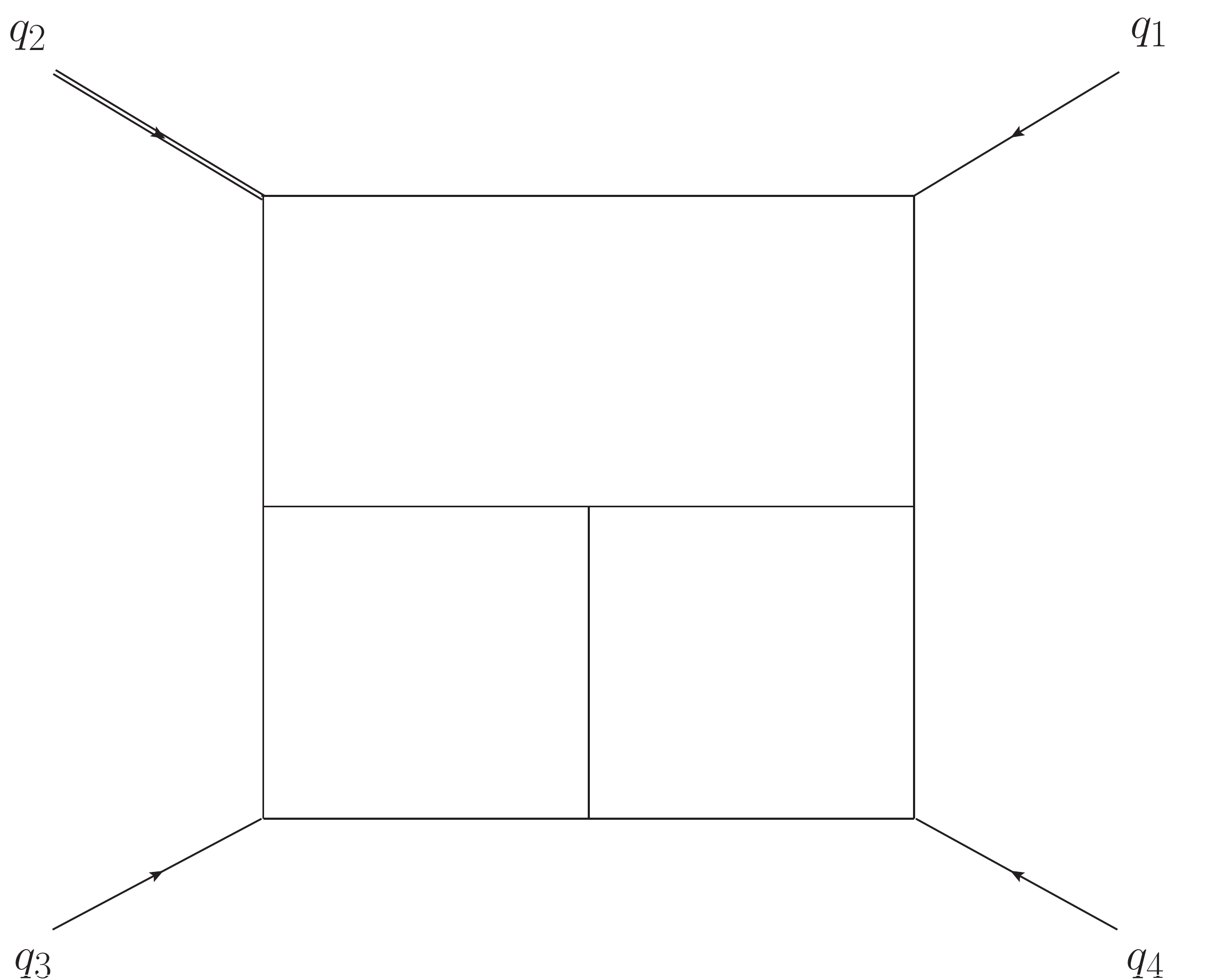}
\includegraphics[width=5.4cm]{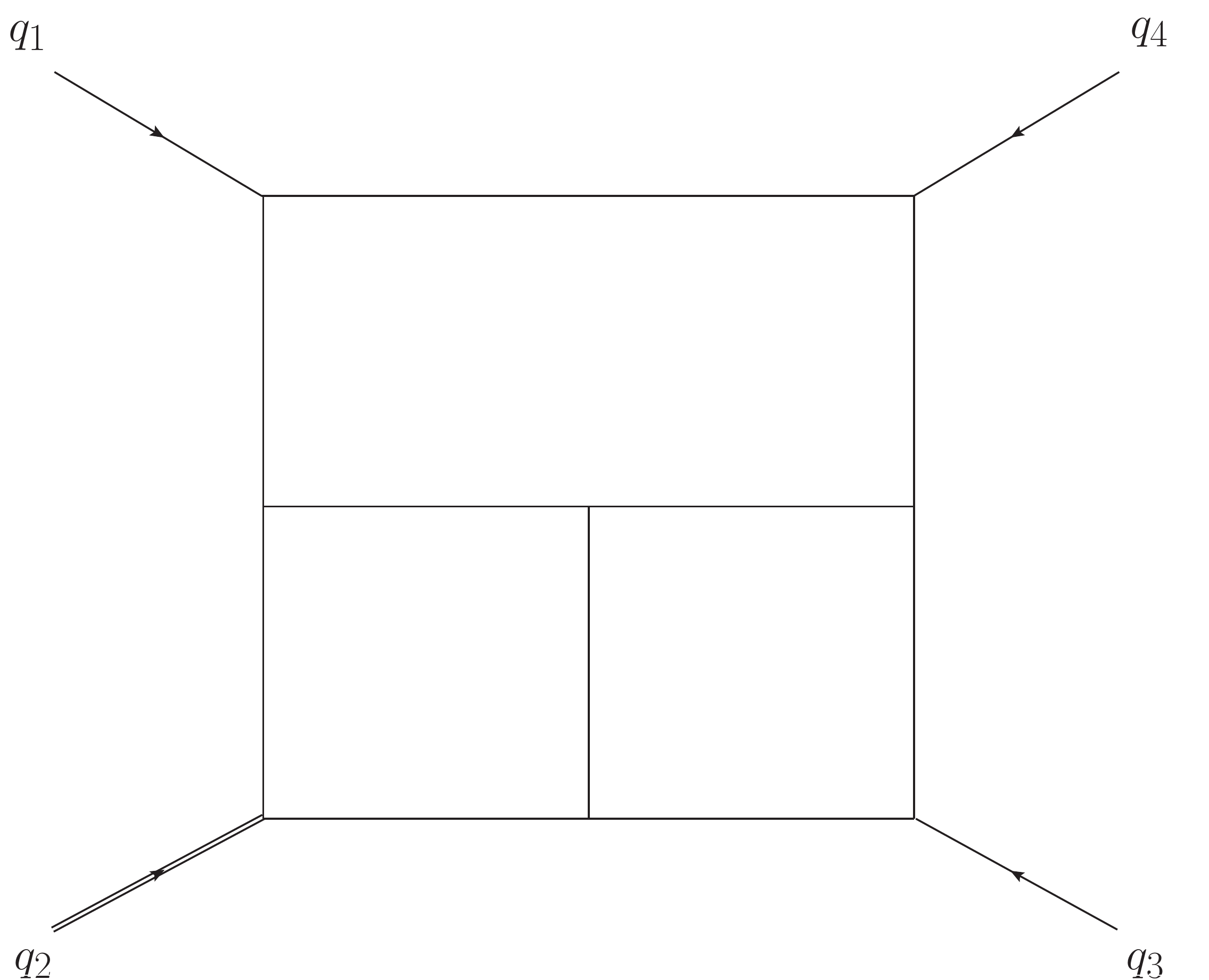}
\caption{The F1 (top), F2 (bottom left) and F3 (bottom right) top-sector diagrams. The double line represents the massive particle and all external momenta are taken to be incoming.}
\label{fig:topgraphs}
\end{figure}

We begin this section by defining the integral families that will be treated in this paper. Even though the ladder-box topology was calculated with our approach in \cite{Canko:2020gqp}, we include it here again, since we will present results for this family as well in physical kinematics. For the sake of brevity we will call F1 the ladder-box topology and F2, F3 the two tennis-court ones, as depicted in figure \ref{fig:topgraphs}.

The corresponding FI are defined through\footnote{Where we use the abbreviation $q_{12}=q_1+q_2$ and $q_{123}=q_1+q_2+q_3$.} 
\begin{align}\label{eq:F1}
    G^{F1}_{a_1\cdots a_{15}}:=& \int \bigg(\prod_{l=1}^3 e^{\gamma_E \epsilon} \frac{d^dk_l}{i\pi^{d/2}}\bigg)\frac{\left(k_1+q_{123}\right){}^{-2
   a_{11}}\left(k_2+q_{123}\right){}^{-2 a_{12}}}{\left(k_1+q_{12}\right){}^{2 a_1}\left(k_2+q_{12}\right){}^{2 a_2}\left(k_3+q_{12}\right){}^{2 a_3}\left(k_1-k_2\right){}^{2 a_4}}\nonumber\\
    &\times \frac{\left(k_2+q_1\right){}^{-2 a_{13}}\left(k_3+q_1\right){}^{-2 a_{14}}\left(k_1-k_3\right){}^{-2 a_{15}}}{\left(k_2-k_3\right){}^{2
   a_5}\left(k_3+q_{123}\right){}^{2 a_6}~k_1^{2 a_7}~k_2^{2 a_8}~k_3^{2 a_9}\left(k_1+q_1\right){}^{2 a_{10}}}
\end{align}

\begin{align}\label{eq:F2}
    G^{F2}_{a_1\cdots a_{15}}:=& \int \bigg(\prod_{l=1}^3 e^{\gamma_E \epsilon} \frac{d^dk_l}{i\pi^{d/2}}\bigg)\frac{\left(k_1+q_{123}\right){}^{-2 a_{11}}~k_2^{-2
   a_{12}}}{\left(k_1+q_{12}\right){}^{2
   a_1}\left(k_2+q_{12}\right){}^{2
   a_2}\left(k_2+q_{123}\right){}^{2
   a_3}}\nonumber\\
   &\times \frac{\left(k_2+q_1\right){}^{-2
   a_{13}}\left(k_3+q_1\right){}^{-2
   a_{14}}\left(k_3+q_{12}\right){}^{-2 a_{15}}}{\left(k_3+q_{123}\right){}^{2 a_4}~k_3^{2
   a_5}~k_1^{2 a_6}\left(k_1+q_1\right){}^{2
   a_7}\left(k_1-k_2\right){}^{2
   a_8}\left(k_1-k_3\right){}^{2
   a_9}\left(k_3-k_2\right){}^{2
   a_{10}}}
\end{align}

\begin{align}\label{eq:F3}
    G^{F3}_{a_1\cdots a_{15}}:=& \int \bigg(\prod_{l=1}^3 e^{\gamma_E \epsilon} \frac{d^dk_l}{i\pi^{d/2}}\bigg)\frac{\left(k_1+q_{12}\right){}^{-2 a_{11}}~k_2^{-2
   a_{12}}}{\left(k_1+q_1\right){}^{2
   a_1}\left(k_2+q_1\right){}^{2
   a_2}\left(k_2+q_{12}\right){}^{2
   a_3}\left(k_3+q_{12}\right){}^{2
   a_4}}\nonumber\\
   &\times\frac{\left(k_2+q_{123}\right){}^{-2 a_{13}}~k_3^{-2
   a_{14}}\left(k_3+q_1\right){}^{-2 a_{15}}}{\left(k_3+q_{123}\right){}^{2
   a_5}\left(k_1+q_{123}\right){}^{2 a_6}~k_1^{2
   a_7}\left(k_1-k_2\right){}^{2
   a_8}\left(k_1-k_3\right){}^{2
   a_9}\left(k_3-k_2\right){}^{2
   a_{10}}}
\end{align}
with $a_i$ being integers and $a_i\leq 0$ for $i=11,\ldots,15$. The external momenta of the considered families obey the following kinematics: $\sum_{i=1}^4 q_i=0,~q_2^2=m^2,~q_i^2=0$ for $i=1,3,4$ and $S_{12}=(q_1+q_2)^2,~S_{23}=(q_2+q_3)^2,~S_{13}=m^2-S_{12}-S_{23}$.

In this paper we employ the SDE approach \cite{Papadopoulos:2014lla} for the analytic computation of families F2 and F3. To do so we parametrise the external momenta by introducing a dimensionless parameter $x$ in the following manner
\begin{equation}\label{eq:xparam}
    q_1 = x p_1, \quad q_2= p_1 + p_2 - x p_1, \quad q_3 = p_3, \quad q_4 = p_4
\end{equation}
where the new momenta $p_i$ are all massless. This parametrisation produces the following mapping for the kinematic invariants between the two momentum configurations
\begin{equation}\label{eq:invmap}
    S_{12} = s_{12}, \quad S_{23} = s_{23} x,\quad m^2 = s_{12} (1-x)
\end{equation}
with $s_{12}=(p_1+p_2)^2,~s_{23}=(p_2+p_3)^2$. The SDE approach \cite{Papadopoulos:2014lla} is an attempt to simplify and systemize, as much as possible, the derivation of the appropriate system of DE satisfied by the MI. Taking $p_i$ to be massless, the introduction of the $x$ parameter in \eqref{eq:xparam} captures the off-shellness of the massive momentum $q_2$. The MI are now dependent on $x$ through the external momenta. This allows us to derive DE by differentiating with respect to $x$. The benefit of this approach is that, when the poles of the DE are rational functions of $x$, the integration of the DE naturally captures the expressibility of MI in terms of GPLs and more importantly makes the problem independent of the number of kinematic scales involved. Note that as $x \rightarrow 1$, the external momentum $q_2$ becomes massless, which corresponds to a simpler loop integral family with one scale less.

\subsection{Scattering kinematics}
When considering the analytic solution of multiloop FI in dimensional regularisation, one usually solves these integrals in the Euclidean region, where all FI are free of branch cuts, and then analytically continues the results to the physical regions of phase-space. This is the approach that we will follow as well. 

At first, by studying the second Symanzik polynomial \cite{Bogner:2010kv} for the top-sector integrals\footnote{By top-sector we mean the integrals with $a_i=1$ for $i=1,\ldots 10$ and $a_i=0$ for $i=11,\ldots15$ in \eqref{eq:F1}, \eqref{eq:F2} and \eqref{eq:F3}.} of families F1, F2 and F3 in their Feynman parameter representation, we can identify the Euclidean region in terms of the kinematic variables $S_{12},~S_{23},~m^2$ such that
\begin{equation}
    S_{12}<0,\quad S_{23}<0,\quad m^2<0.
\end{equation}
For scattering kinematics we have three physical regions when considering $2\to2$ processes with one massive particle. For convenience we will denote them as the $s,~t~\text{and}~u$ channels appropriately:
\begin{align}
    \text{s-channel}:&~  m^2>0,\quad S_{12}\geq m^2,\quad S_{23}\leq0,\quad S_{13}\leq0 \\
    \text{t-channel}:&~  m^2>0,\quad S_{12}\leq 0,\quad S_{23}\geq m^2,\quad S_{13}\leq0 \\
    \text{u-channel}:&~  m^2>0,\quad S_{12}\leq 0,\quad S_{23}\leq0,\quad S_{13}\geq m^2.
\end{align}

Since we will use the SDE approach for the solution of \eqref{eq:F2} and \eqref{eq:F3}, we would like to have the corresponding limits for each region of phase-space expressed in terms of the $x,~s_{12},~s_{23}$ variables. The mapping of \eqref{eq:invmap} allows us to do so, although for reasons that will become clear at a later stage, we define the ratio $y=\frac{s_{23}}{s_{12}}$ and use the variables $x,~y,~s_{12}$. Our approach therefore will be to compute all MI in terms of real-valued GPLs in the Euclidean region
\begin{equation}\label{eq:ERx}
    0<x<1,\quad s_{12}<0,\quad 0<y<1
\end{equation}
and then, using tools such as \texttt{HyperInt}\cite{Panzer:2014caa} and \texttt{PolyLogTools}\cite{Duhr:2019tlz}, analytically continue our solutions in the physical regions
\begin{align}\label{eq:PRx}
    \text{s-channel}:&~  0<x<1,\quad s_{12}>0,\quad -1\leq y\leq0 \\
    \text{t-channel}:&~  1<x,\quad s_{12}<0,\quad y\leq -1 \\
    \text{u-channel}:&~  1<x,\quad s_{12}<0,\quad y\geq 0.
\end{align}

\section{Canonical differential equations}
\label{sec:cde}
 Using IBP tools such as \texttt{FIRE6} \cite{Smirnov:2019qkx} and \texttt{Kira2} \cite{Klappert:2020nbg} we identified a minimal set of 117 MI (3 at the top-sector) for family F2 and 166 MI (4 at the top sector) for family F3. Although today exist several automated packages for the construction of a pure basis which satisfies a canonical DE \cite{Henn:2020lye, Meyer:2017joq, Prausa:2017ltv, Gituliar:2017vzm, Dlapa:2020cwj}, finding a pure basis for a generic integral family in practise is still a non-trivial task, assuming that such a basis even exists. 
 
 In order to construct pure bases for families F2 and F3 we used a combined approach, attacking the problem with the following techniques. For several low sectors, involving up to seven propagators, we used the approach based on Magnus series expansions \cite{Argeri:2014qva}, which was already successfully applied for the construction of a pure basis for family F1 in \cite{DiVita:2014pza}. For higher sectors, involving up to nine propagators, we used the \texttt{Mathematica} package \texttt{DlogBasis} \cite{Henn:2020lye} to identify appropriate candidates as pure basis elements. The last but most extensively used approach is a heuristic method based on \cite{Wasser:2018qvj}, working loop-by-loop and using already known one, two- and three-loop pure basis elements \cite{DiVita:2014pza, Henn:2014lfa, Dlapa:2021qsl}. For a more detailed discussion we refer the interested reader to \cite{Canko:2021hvh}.

All approaches mentioned above are valid for the determination of a \textit{candidate} pure basis. The ultimate test that such a basis has been obtained is whether it satisfies a canonical DE. Indeed, by differentiating with respect to $x$ we were able to obtain the following SDE in canonical form for families F2 and F3,
\begin{equation}\label{eq:cansde}
\partial_{x} \textbf{g}=\epsilon \left( \sum_{i=1}^{4} \frac{\textbf{M}_i}{x-l_i} \right) \textbf{g}
\end{equation}
where $\textbf{g}$ is the pure basis and $\textbf{M}_i$ are the residue matrices corresponding to each letter $l_i$. All kinematic dependence is included in the letters $l_i$, leaving the matrices $\textbf{M}_i$ to consist solely of rational numbers. We have found an alphabet consisting of the four following letters
\begin{equation}\label{eq:alphabetx}
 l_1=0,~l_2=1,~l_3=\frac{1}{1+y},~l_4=-\frac{1}{y}.
\end{equation}
Notice that here we follow \cite{Papadopoulos:2015jft,Canko:2020gqp} for the definition of the letters of the alphabet, which is different from the standard notation \cite{Duhr:2011zq, Duhr:2012fh, Duhr:2014woa}. Usually the so-called $\mathrm{d}\log$ form of a system of canonical differential equations is given as $\mathrm{d}\textbf{g}(\vec{x}, \epsilon) = \epsilon \left(\sum_{i} \textbf{M}_i \mathrm{d}\log W_i (\vec{x}) \right) \textbf{g}(\vec{x}, \epsilon)$, where the alphabet $W_i (\vec{x})$ is in terms of rational or algebraic functions of the independent variables. The standard $\mathrm{d}\log$ form is equivalent to \eqref{eq:cansde} for $W_i (\vec{x}) = x-l_i$.

A few comments here are in order. Firstly, the same alphabet was found for the F1 family in \cite{Canko:2020gqp}. Secondly, we have observed that in every case where both the standard approach, i.e. differentiating with respect to all kinematic invariants, and the SDE approach have been applied, we have found a reduced number of logarithmic singularities in the canonical SDE. For the ladder topology, in \cite{DiVita:2014pza} the corresponding canonical DE was characterised by six logarithmic singularities, i.e. six $W_i (\vec{x})$ functions according to the standard $\mathrm{d}\log$ notation. With the SDE approach, we obtain a canonical DE in $x$ characterised by four $W_i (\vec{x})$ functions. This reduction in the number of poles of the corresponding canonical DE stems from the fact that we only consider the differentiation with respect to $x$, leaving the $s_{12}$- and $y$-dependence of the MI to be explicitly treated in the boundary terms. Furthermore, the fact that all three-loop planar families with one off-shell leg are characterised by the same alphabet is also an interesting one. This was also observed in the fully massless case, both for the planar \cite{Henn:2013tua} and the non-planar \cite{Henn:2020lye} topologies. If this property holds true for the three-loop non-planar families with one off-shell leg as well, then we expect that their solution in terms of GPLs will be an achievable goal in the foreseeable future. 

The simplicity of the alphabet \eqref{eq:alphabetx} in $x$ allows for a straightforward solution of \eqref{eq:cansde} in terms of GPLs. The solution can be written in the following compact form up to weight six:
\begin{equation}
\label{eq:solution}
{\footnotesize
\begin{split}
\textbf{g}&=\epsilon^0 \textbf{b}_0^{(0)}+\epsilon \left(\sum {\cal G}_i \textbf{M}_i \textbf{b}_0^{(0)}+\textbf{b}_0^{(1)}\right)+\epsilon^2 \left(\sum {\cal G}_{ij} \textbf{M}_i\textbf{M}_j\textbf{b}_0^{(0)}+\sum {\cal G}_i \textbf{M}_i \textbf{b}_0^{(1)}+\textbf{b}_0^{(2)} \right)+ \dots \\  
&+ \epsilon^6 \left(\textbf{b}_0^{(6)}+ \sum {\cal G}_{ijklmn} \textbf{M}_i \textbf{M}_j \textbf{M}_k \textbf{M}_l \textbf{M}_m \textbf{M}_n \textbf{b}_0^{(0)} + \sum {\cal G}_{ijklm} \textbf{M}_i \textbf{M}_j \textbf{M}_k \textbf{M}_l \textbf{M}_m \textbf{b}_0^{(1)} \right. \\
&+\left. \sum {\cal G}_{ijkl} \textbf{M}_i \textbf{M}_j \textbf{M}_k \textbf{M}_l \textbf{b}_0^{(2)} +\sum {\cal G}_{ijk} \textbf{M}_i\textbf{M}_j\text{M}_k \textbf{b}_0^{(3)}+\sum {\cal G}_{ij} \textbf{M}_i\textbf{M}_j\textbf{b}_0^{(4)}+\sum {\cal G}_i \textbf{M}_i \textbf{b}_0^{(5)} \right)
\end{split}}
\end{equation}
were $\mathcal{G}_{ab\ldots}:= \mathcal{G}(l_a,l_b,\ldots;x)$ represent the GPL. The $\textbf{b}_0^{(i)}$ terms represent the boundary terms that need to be determined, with $i$ indicating the corresponding weight, and consist of Zeta functions $\zeta(i)$ and logarithms $\{\log(-s_{12}),~\log(y)\}$ of weight $i$. Our results are presented in such a way that each coefficient of $\epsilon^i$ has transcendental weight $i$. If we assign weight $-1$ to $\epsilon$, then \eqref{eq:solution} has uniform weight zero.

\subsection{Boundary terms}
In order to determine the necessary boundary terms $\textbf{b}_0^{(i)}$ in \eqref{eq:solution} we will employ the techniques developed in \cite{Canko:2020gqp, Canko:2020ylt}, as well as showcase a new way of using them that significantly simplifies the entire procedure.

In general we need to calculate the $x\to0$ limit of each pure basis element. Our first step is to exploit the canonical SDE at the limit $x\to0$ and define through it the resummation matrix 
\begin{equation}\label{eq:resmzero}
    \textbf{R} =\textbf{S} \mathrm{e}^{\epsilon \textbf{D} \log(x)} \textbf{S}^{-1}
\end{equation}
where the matrices $\textbf{S},~ \textbf{D}$ are obtained through the Jordan decomposition\footnote{For an earlier use of the Jordan decomposition method see \cite{Dulat:2014mda, Mastrolia:2017pfy}.} of the residue matrix for the letter $l_1=0$,  $\textbf{M}_1$,
\begin{equation}\label{eq:jordandeczero}
    \textbf{M}_1 =\textbf{S} \textbf{D} \textbf{S}^{-1}.
\end{equation}
On the other hand, through IBP reduction, the elements of the pure basis can be related to a set of FI $\textbf{G}$,
\begin{equation}\label{eq:gtomasters}
    \textbf{g}=\textbf{T}\textbf{G}.
\end{equation}
Furthermore using the expansion by regions method \cite{Jantzen:2012mw} as implemented in the \texttt{asy} code which is shipped along with \texttt{FIESTA4} \cite{Smirnov:2015mct}, we can obtain information for the asymptotic behaviour of the FI in terms of which we express the pure basis of MI \eqref{eq:gtomasters} in the limit $x\to0$,
\begin{equation}\label{eq:regions}
    {G_i}\mathop  = \limits_{x \to 0} \sum\limits_j x^{b_j + a_j \epsilon }G^{(b_j + a_j \epsilon)}_{i} 
\end{equation}
where $a_j$ and $b_j$ are integers and $G_i$ are the individual members of the basis $\textbf{G}$ of FI in \eqref{eq:gtomasters}. This analysis allows us to construct the following relation
\begin{equation}\label{eq:bounds}
    \mathbf{R} \mathbf{b}=\left.\lim _{x \rightarrow 0} \mathbf{T} \mathbf{G}\right|_{\mathcal{O}\left(x^{0+a_{j} \epsilon}\right)}
\end{equation}
where the right-hand side implies that, apart from the terms $x^{a_i  \epsilon}$ coming from \eqref{eq:regions}, we expand around $x=0$, keeping only terms of order $x^0$. Equation \eqref{eq:bounds} allows us in principle to determine all boundary constants $\textbf{b}=\sum_{i=0}^{6}~ \epsilon^i~ b_0^{(i)}$. 

More specifically, equation \eqref{eq:bounds} produces two kinds of relations. The first one, called \textit{pure relations} in \cite{Canko:2020gqp}, is in the form of linear relations with numerical rational coefficients between boundary terms $\textbf{b}$ and the second one is in the form of linear relations that contain boundary terms $\textbf{b}$ and region-integrals $G^{(b_j + a_j \epsilon)}_{i}$ in the Feynman parameter representation. These relations have been sufficient in past studies \cite{Canko:2020gqp, Canko:2020ylt, Syrrakos:2020kba, Syrrakos:2021nij} for the calculation of all boundary terms. Their application for families F2 and F3 however requires the computation of several highly non-trivial region-integrals.

To circumvent this issue, we take advantage of the fact that the $x\to1$ limit of our solution \eqref{eq:solution} will yield the result for the fully massless integral family \cite{Canko:2020gqp}, which is already available in \cite{Henn:2013tua}. Our approach therefore is to exploit the pure relations between boundary terms $\textbf{b}$, compute a minimal number of region-integrals and then construct the full solution by explicitly keeping the undetermined boundary terms in the final result. Then we extract the $x\to1$ limit using the techniques of \cite{Canko:2020gqp} and use the known analytic results of \cite{Henn:2013tua} to fix all remaining unknown boundaries.

To exemplify this approach we will present here the computation of all boundary terms for the F3 family, assuming we have already solved family F2. Starting with the construction of equation \eqref{eq:bounds}, we determine the so-called \textit{pure relations}. These relations usually determine the boundary terms for the top-sector basis elements, which are the most non-trivial to compute since they involve the most complicated region-integrals. For the case of family F3 these \textit{pure relations} for the four top-sector basis elements are
{\small
\begin{align}
    b_{163}=&-\frac{371 b_1}{132}-\frac{8 b_2}{11}+\frac{12 b_4}{11}-\frac{637 b_5}{33}+32
   b_{10}+\frac{64 b_{11}}{3}+\frac{37 b_{12}}{6}-15 b_{13}+\frac{3 b_{15}}{22}-15
   b_{18}\nonumber\\
   &+12 b_{19}-\frac{9 b_{22}}{4}-\frac{9 b_{23}}{2}+3 b_{24}-\frac{15
   b_{25}}{22}-\frac{b_{26}}{22}+12 b_{29}+8 b_{30}-6 b_{31}+\frac{75 b_{35}}{11}+6
   b_{36}\\
   &-\frac{84 b_{38}}{11}+9 b_{43}+36 b_{52}-18 b_{53}-21 b_{54}+6 b_{57}-\frac{6
   b_{77}}{11}-b_{81}-12 b_{87}+3 b_{88}-2 b_{123}\nonumber\\
   &-6 b_{130}+2 b_{137}+6 b_{144}-2
   b_{152}-2 b_{159},\nonumber \\
   b_{164}=&-\frac{745 b_1}{264}-\frac{13 b_2}{66}+\frac{6 b_4}{11}-\frac{359
   b_5}{22}+\frac{92 b_{10}}{3}+\frac{52 b_{11}}{3}+\frac{31 b_{12}}{6}-\frac{89
   b_{13}}{6}+\frac{51 b_{15}}{22}\nonumber\\
   &-\frac{97 b_{18}}{6}+12 b_{19}-\frac{9
   b_{22}}{8}-\frac{9 b_{23}}{2}+3 b_{24}-\frac{15 b_{25}}{44}+\frac{5 b_{26}}{22}+10
   b_{29}+\frac{20 b_{30}}{3}-6 b_{31}\\
   &+\frac{75 b_{35}}{22}+5 b_{36}-\frac{42
   b_{38}}{11}+9 b_{43}+36 b_{52}-18 b_{53}-\frac{33 b_{54}}{2}+6 b_{57}-\frac{3
   b_{77}}{11}-b_{81}\nonumber\\
   &-12 b_{87}+3 b_{88}-2 b_{123}-6 b_{130}+2 b_{137}+4 b_{144}-3
   b_{152}-3 b_{159},\nonumber\\
   b_{165}=&0,\\
   b_{166}=&-\frac{1531 b_1}{4752}-\frac{128 b_2}{297}+\frac{47 b_4}{33}-\frac{1891
   b_5}{396}+\frac{74 b_{10}}{9}+\frac{20 b_{11}}{3}+\frac{7 b_{12}}{3}-\frac{127
   b_{13}}{36}-\frac{415 b_{15}}{264}\nonumber\\
   &+\frac{13 b_{16}}{8}+\frac{10 b_{17}}{3}-\frac{47
   b_{18}}{36}-2 b_{19}+\frac{5 b_{20}}{6}-\frac{21 b_{22}}{16}-\frac{11
   b_{23}}{6}+\frac{5 b_{24}}{12}-\frac{35 b_{25}}{132}-\frac{6 b_{26}}{11}\nonumber\\
   &+\frac{16
   b_{29}}{3}+\frac{32 b_{30}}{9}-\frac{10 b_{31}}{3}+\frac{581 b_{35}}{132}+\frac{29
   b_{36}}{18}-\frac{197 b_{38}}{33}+\frac{3 b_{43}}{2}-\frac{14 b_{49}}{3}+7 b_{52}\\
   &-5
   b_{53}-\frac{89 b_{54}}{12}+\frac{13 b_{57}}{3}-\frac{8 b_{60}}{3}-\frac{b_{61}}{6}+2
   b_{62}-\frac{7 b_{77}}{33}-\frac{b_{81}}{6}+3 b_{83}-\frac{b_{84}}{2}-\frac{13
   b_{87}}{6}\nonumber\\
   &+\frac{7 b_{88}}{12}-\frac{2 b_{89}}{3}+\frac{5
   b_{97}}{6}-\frac{b_{108}}{3}-\frac{2 b_{123}}{3}-b_{130}+\frac{2 b_{137}}{3}+2
   b_{144}-\frac{4 b_{152}}{3}-\frac{2 b_{159}}{3}\nonumber.
\end{align}}Similar relations are obtained for 109 basis elements in total, leaving the following basis elements undetermined,
\begin{align}
    \big\{&b_1,b_2,b_4,b_5,b_7,b_{10},b_{11},b_{12},b_{13},b_{15},b_{16},b_{17},b_{18},b_{19}
   ,b_{20},b_{22},b_{23},b_{24},b_{25},b_{26},b_{29},b_{30},b_{31},b_{35},\nonumber\\
   &b_{36},b_{38},b
   _{43},b_{49},b_{50},b_{52},b_{53},b_{54},b_{57},b_{59},b_{60},b_{61},b_{62},b_{77},b_{
   81},b_{83},b_{84},b_{85},b_{87},b_{88},b_{89},b_{97},b_{108},\nonumber\\
   &b_{116},b_{123},b_{130},b
   _{135},b_{137},b_{144},b_{151},b_{152},b_{157},b_{159}\big\}.
\end{align}
From the above list of boundary terms, most of them are known either from family F1 or from family F2. The only genuinely unknown boundary terms therefore are
\begin{equation}
    \big\{b_{108},b_{123},b_{135},b_{144},b_{157},b_{159}\big\}.
\end{equation}
Furthermore, $b_{108}$ can be easily calculated through a direct integration of the region-integrals that appear in its result coming from \eqref{eq:bounds},
\begin{align}
    b_{108}=&-2 b_{19}+\frac{3 b_{21}}{4}+s_{12}^2 \epsilon ^5 G_{111101012000000}^{(-2 \epsilon )}+4 s_{12} \epsilon^4 G_{1022010110-10000}^{(-\epsilon )}\nonumber\\
   &-3 s_{12}^2 \epsilon ^4
   G_{112201001000000}^{(-\epsilon )}+6 s_{12}
   \epsilon ^5 G_{0 1 1 1 0 1 0 1 2 0 0 0 0 0 0}^{(0)}.
\end{align}
As you can see, the boundary term $b_{108}$ involves four region-integrals with the most non-trivial having seven propagators, i.e requiring the integration of seven Feynman parameters - in general a non-trivial, but in this case achievable task. This leaves us with the following five undetermined boundary terms $\big\{b_{123},b_{135},b_{144},b_{157},b_{159}\big\}$. If we were to continue with this approach for the rest of the undetermined boundary terms, we would have to compute several highly non-trivial region-integrals, involving the direct integration of up to nine Feynman parameters. Such calculations can be extremely difficult, even if the result of the integration is rather simple. 

To obtain the remaining boundary terms, we will exploit a feature of the SDE approach which is that by taking the $x\to1$ limit of our results, we recover the solution for the massless momentum configuration, as can be seen from \eqref{eq:xparam} and \eqref{eq:invmap}. Our approach involves the following steps:
\begin{enumerate}
    \item Construct solution \eqref{eq:solution} using a simple ansatz for the undetermined boundary terms, i.e. $b_i = \sum_{k=0}^6 a(i,k) \epsilon^k$.
    \item Construct a resummation matrix from the $\textbf{M}_2$ residue matrix and using the shuffle properties of GPLs extract the part of the solution \eqref{eq:solution} which is regular at $x=1$. These two ingredients allow us to obtain the $x\to1$ limit of \eqref{eq:solution} \cite{Canko:2020gqp}.
    \item Map the $x\to1$ limit of \eqref{eq:F3}, i.e. the massless tennis-court, to the same family defined in \cite{Henn:2013tua}.
    \item Map the pure basis for the massless tennis-court of \cite{Henn:2013tua} to the pure basis that can be obtained from the $x\to1$ limit of the pure basis of F3 using the techniques of \cite{Canko:2020gqp}.
\end{enumerate}
The mapping between our results and that of \cite{Henn:2013tua} yields that the variable $y=\frac{s_{23}}{s_{12}}$ that we introduced earlier is equal to the dimensionless parameter $x$ used therein\footnote{Not to be confused with the SDE parameter $x$ that we use throughout this paper.}. In the case of the F2 family the respective relation is $y=\frac{1}{x}$. This exemplifies the reason we introduced $y$ in the first place. 

The above steps allow us to fix all remaining boundary terms in a purely analytical way. We find that a general ansatz for all boundary terms can be constructed,{\small
\begin{align}
     b_i &= c(i,0) + \epsilon  \log (y) c(i,1) + \epsilon ^2 \left(\frac{1}{2} \log ^2(y) c(i,2,2)+\frac{1}{6} \pi ^2 c(i,2,1)\right)\nonumber\\
    &+\epsilon ^3 \left(\frac{1}{6} \log ^3(y) c(i,3,2)+\frac{1}{6} \pi ^2 \log (y) c(i,3,1)+\zeta (3) c(i,3,3)\right)\nonumber\\
    &+\epsilon ^4 \left(\zeta (3) \log (y) c(i,4,4)+\frac{1}{24} \log ^4(y) c(i,4,3)+\frac{1}{12} \pi ^2 \log ^2(y)
   c(i,4,2)+\frac{1}{90} \pi ^4 c(i,4,1)\right)\nonumber\\
   &+\epsilon ^5 \bigg(\frac{1}{2} \zeta (3) \log ^2(y) c(i,5,5)+\frac{1}{120} \log ^5(y) c(i,5,3)+\frac{1}{36} \pi ^2 \log
   ^3(y) c(i,5,2)\\
   &+\frac{1}{90} \pi ^4 \log (y) c(i,5,1)+\zeta (5) c(i,5,6)+\frac{1}{6} \pi ^2 \zeta (3) c(i,5,4)\bigg)\nonumber\\
   &+\epsilon ^6 \bigg(\frac{1}{6} \zeta (3) \log ^3(y) c(i,6,6)+\log (y) \left(\frac{1}{6} \pi ^2 \zeta (3) c(i,6,5)+\zeta (5)
   c(i,6,8)\right)+\frac{1}{720} \log ^6(y) c(i,6,4)\nonumber\\
   &+\frac{1}{144} \pi ^2 \log ^4(y) c(i,6,3)+\frac{1}{180} \pi ^4 \log
   ^2(y) c(i,6,2)+\zeta (3)^2 c(i,6,7)+\frac{1}{945} \pi ^6 c(i,6,1)\bigg)\nonumber
\end{align}}where we have multiplied by $(-s_{12})^{(3\epsilon)}$ and expanded up to $\epsilon^6$ to make the ansatz more compact. This is of course the most general form that is compatible with the scaling and universal transcendentality of the boundaries. If the same alphabet holds for some or all the non-planar families one could envision obtaining the boundary terms by fitting numerically this ansatz for the remaining families of integrals using the PSLQ algorithm \cite{pslq}.

\section{Results}
\label{sec:results}
\subsection{Analytic continuation}

As already mentioned, our results are expressed in terms of GPLs up to weight six and thus can be numerically computed at high precision using automated tools, like \texttt{GinaC} \cite{Vollinga:2004sn}. The weight W=$1\ldots 6$ is identified as the number of letters $l_i$ in $\mathcal{G}(l_i,\ldots;x)$. For the evaluations to be fast and efficient the GPLs should not contain letters $l_i$ along the integration path connecting the origin and the argument $x$, i.e. $l_i\notin[0,x]$. Although this holds true in the Euclidean region, this is not the case in the physical regions anymore. One can get over this problem by using fibration-basis techniques \cite{Duhr:2019tlz,Panzer:2014caa} for changing appropriately the arguments of GPLs at each physical region and making them real-valued. We followed this procedure using \texttt{HyperInt} \cite{Panzer:2014caa}, aiming to obtain results in terms of real-valued GPLs which can be efficiently evaluated numerically, thus making them  well-suited for phenomenological applications.

\begin{table}[h!]
\begin{adjustwidth}{-1.in}{-1.in} 
\begin{center}
 \begin{tabular}{| c || c | c || c | c |} 
 \hline
 \textbf{Regions} & \textbf{Letters} & \textbf{Argument} & \textbf{Letters} & \textbf{Argument} \\
 \hline
 Euclidean & $\{0,1,-1/y,1/(1+y)\}$ & $x$ & $-$  & $-$\\ 
 \hline
 s-channel & $\{0,1,-1/y,1/(1+y)\}$ & $x$ & $-$  & $-$\\  
 \hline
 t-channel & $\{0,1,1+y,-y\}$ & $1/x$ & $\{0,1\}$ & $-1/y$\\  
 \hline
 u-channel & $\{0,1,1+y,-y\}$ & $1/x$ & $\{0,-1\}$ & $y$\\  
 \hline
\end{tabular}
\end{center}
\end{adjustwidth}
\caption{Structure of GPLs appearing in each of the 4 kinematic regions.}
\label{tab:t1}
\end{table}

In order to make more clear the form of our solutions, we present in table \ref{tab:t1} the arguments and the letters of the GPLs in the Euclidean and each of the physical regions after using \texttt{HyperInt}. Note that prior to the use of fibration-basis techniques, no GPLs with arguments depending on $y$ are present. However, casting each GPL in a fibration basis which contains real-valued GPLs and explicit imaginary terms, produces GPLs with $y$-dependent argument as well.

In table \ref{tab:t2} we provide an analysis regarding the number of GPLs that appear in each transcendental weight, for each kinematic region. In the same table we quote also the total number of GPLs and the timing for their numerical evaluation using \texttt{GinaC}. All the families have the same GPLs at every region except for the F1 family, which in the Euclidean region (and in the s-channel) contains 42 less GPLs, all of them of weight 6. The number of GPLs is the same for the Euclidean and the s-channel due to the fact that the GPLs of the Euclidean region are already real valued in the s-channel and thus their is no need of using fibration-basis techniques for them. 

\begin{table}[h!]
\begin{adjustwidth}{-1.in}{-1.in} 
\begin{center}
 \begin{tabular}{| c | c | c | c | c | c | c | c | c |} 
 \hline
 \textbf{Regions} & \textbf{$W=1$} & \textbf{$W=2$} & \textbf{$W=3$} & \textbf{$W=4$} & \textbf{$W=5$} & \textbf{$W=6$} & \textbf{Total} & \textbf{Timings (sec)} \\
 \hline
 Euclidean & 4 & 14 & 50 & 124 & 367 & 734 & 1293 & 39.0225769\\ 
 \hline
 s-channel & 4 & 14 & 50 & 124 & 367 & 734 & 1293 & 39.2172529\\  
 \hline
 t-channel & 6 & 18 & 58 & 155 & 419 & 603 & 1259 & 62.0567800\\  
 \hline
 u-channel & 5 & 16 & 54 & 147 & 403 & 572 & 1197 & 55.1049640\\  
 \hline
\end{tabular}
\end{center}
\end{adjustwidth}
\caption{Number of GPLs per transcendental weight and per kinematic region, and timings for the numerical evaluation of the total GPLs. The quoted timings are obtained using the \texttt{Ginsh} command of \texttt{PolyLogTools}, running 1-core in a personal laptop (i7 processor, 8-core, 16GB RAM). The phase-space points that were used are \eqref{eupoint} for the Euclidean region and \eqref{spoint}, \eqref{tpoint}, \eqref{upoint} for the correspoing physical regions.}
\label{tab:t2}
\end{table}

\subsection{Numerical checks}

For the validation of our results we have performed various numerical checks in the Euclidean and physical regions. More specifically, for the Euclidean region we compared numerically every MI of each family with \texttt{pySecDec} \cite{Borowka:2017idc} and \texttt{Fiesta} \cite{Smirnov:2015mct} for the point
\begin{equation}
\label{eupoint}
s_{12} \rightarrow -7, \, y \rightarrow 3/7, \, x \rightarrow 1/4
\end{equation}
finding perfect agreement within the numerical accuracy provided by these programs. 

Regarding our results for physical regions, obtaining numerical results for all MI  using \texttt{pySecDec} or \texttt{Fiesta} can be rather challenging, and in some cases even impossible. To that end, in order to check that the analytic continuation was correct, we determined a specific set of low-sector MI (with up to 7 propagators), which contained the total set of GPLs at each region and we numerically compared them with \texttt{pySecDec} and \texttt{Fiesta} for the points 
\begin{align}
    &\text{s-channel}:~  s_{12} \rightarrow 2, \, y \rightarrow -1/2, \, x \rightarrow 1/4 \label{spoint}\\
    &\text{t-channel}:~  s_{12} \rightarrow -2, \, y \rightarrow -3/2, \, x \rightarrow 5/3 \label{tpoint}\\
    &\text{u-channel}:~  s_{12} \rightarrow -2, \, y \rightarrow 3/2, \, x \rightarrow 5/3.\label{upoint}
\end{align}
After properly choosing the branch of the analytic continuation, i.e. fixing the values of the auxiliary functions $\{\delta(1/x), \delta(-1/y), \delta(y)\}$ coming from \texttt{HyperInt}, perfect agreement was found for every MI that we checked.

\subsection{Analytic checks}
For all MI of families F2 and F3 we performed analytic checks at the limit $x\to1$ against the results of \cite{Henn:2013tua} in the Euclidean region. Even though this limit was used to obtain some of the necessary boundary terms, ensuring that all MI of families F2 and F3 correctly reproduce the results of \cite{Henn:2013tua} at the limit $x\to1$ constitutes a non-trivial check. 

\subsection{Ancillary files}
Several of our results are available in \cite{gitfiles} in \texttt{Mathematica} readable format. More specifically, we include the following files:
\begin{itemize}
    \item The file \texttt{input.m} contains the propagators from \eqref{eq:F1}, \eqref{eq:F2}, \eqref{eq:F3} as well as their kinematics, the mapping \eqref{eq:invmap} and the alphabet \eqref{eq:alphabetx} which is the same for all three integral families.
    \item The directories Fi, with $i=1,2,3$, which contain:
    \begin{itemize}
        \item \texttt{Fibasis.m}: the pure basis.
        \item \texttt{FiDE.m}: the canonical DE.
        \item \texttt{FiBounds.m}: boundary terms for solving the canonical DE in the Euclidean region.
        \item \texttt{FiE.m}: the solution in the Euclidean region.
        \item \texttt{Fis.m}: the solution in the s-channel.
        \item \texttt{Fit.m}: the solution in the t-channel.
        \item \texttt{Fiu.m}: the solution in the u-channel.
    \end{itemize}
\end{itemize}
We also provide the files \texttt{Fibasis.m}, \texttt{FiDE.m} and \texttt{FiBounds.m} with the \texttt{arXiv} submission of our paper.

\section{Conclusion}
\label{sec:conclusion}

In this paper we presented the analytic computation of the remaining planar three-loop tennis-court families for $2\to2$ processes involving massless propagators and one off-shell leg, in terms of GPLs of up to transcendental weight six. We also presented results for all three-loop planar families, the so-called ladder-box topology with one off-shell leg as well as the two tennis-courts, for scattering kinematics in terms of real-valued polylogarithmic functions, making our results well-suited for phenomenological studies of processes such as vector boson decaying to 3-jets or $gg\to H + jet$ in gluon fusion, at N3LO precision. 

We constructed pure bases for the two tennis-court families and derived canonical DE in the SDE approach, obtaining a simple four-letter alphabet \eqref{eq:alphabetx}, which was identical to the one found for the ladder-box topology in \cite{Canko:2020gqp}. It would be interesting to apply the standard DE approach to the pure bases of the two tennis-court families, i.e. differentiate with respect to all kinematic invariants, and see if their alphabets match the one of the ladder-box topology obtained in \cite{DiVita:2014pza}.


A natural next step is the consideration of the non-planar three-loop families with one massive leg. For the massless case \cite{Henn:2020lye}, all integral families satisfied canonical DE with the same alphabet as in the planar families \cite{Henn:2013tua}. It will be interesting to see if the non-planar families with one massive leg exhibit the same behaviour as well. Constructing pure bases for all of them is a challenging, yet we believe not insurmountable task. 

The mathematical simplicity of the canonical DE \eqref{eq:cansde} for all planar families was instrumental in obtaining results in terms of real-valued GPLs for all physical regions, using existing tools \cite{Panzer:2014caa, Duhr:2019tlz} for their analytic continuation. Should such simplicity be found for the non-planar families, we are confident that their solution in terms of real-valued GPLs for all physical regions of phase-space would be an achievable goal. 

\acknowledgments

All figures have been drawn using \texttt{JaxoDraw} \cite{Binosi:2008ig}. We would like to thank Federico Gasparotto and Luca Mattiazzi for their help with the Magnus series expansions method at the first stages of this project and for many fruitful discussions. The research work of DC was supported by the Hellenic Foundation for Research and Innovation (HFRI) under the HFRI Ph.D. Fellowship grant (Fellowship Number: 554). NS was supported by the Excellence Cluster ORIGINS funded by the Deutsche Forschungsgemeinschaft (DFG, German Research Foundation) under Germany's Excellence Strategy - EXC-2094 - 390783311.

\appendix

\section{Adjacency conditions in the SDE approach}
\label{sec:appA}
In this section we comment on the interpretation of the adjacency conditions discovered in \cite{Dixon:2020bbt, Chicherin:2020umh}, within the SDE approach. In order to do this, firstly, we make a connection between the standard DE method and the SDE approach.

\begin{figure} [h!]
\centering
\includegraphics[width=9cm]{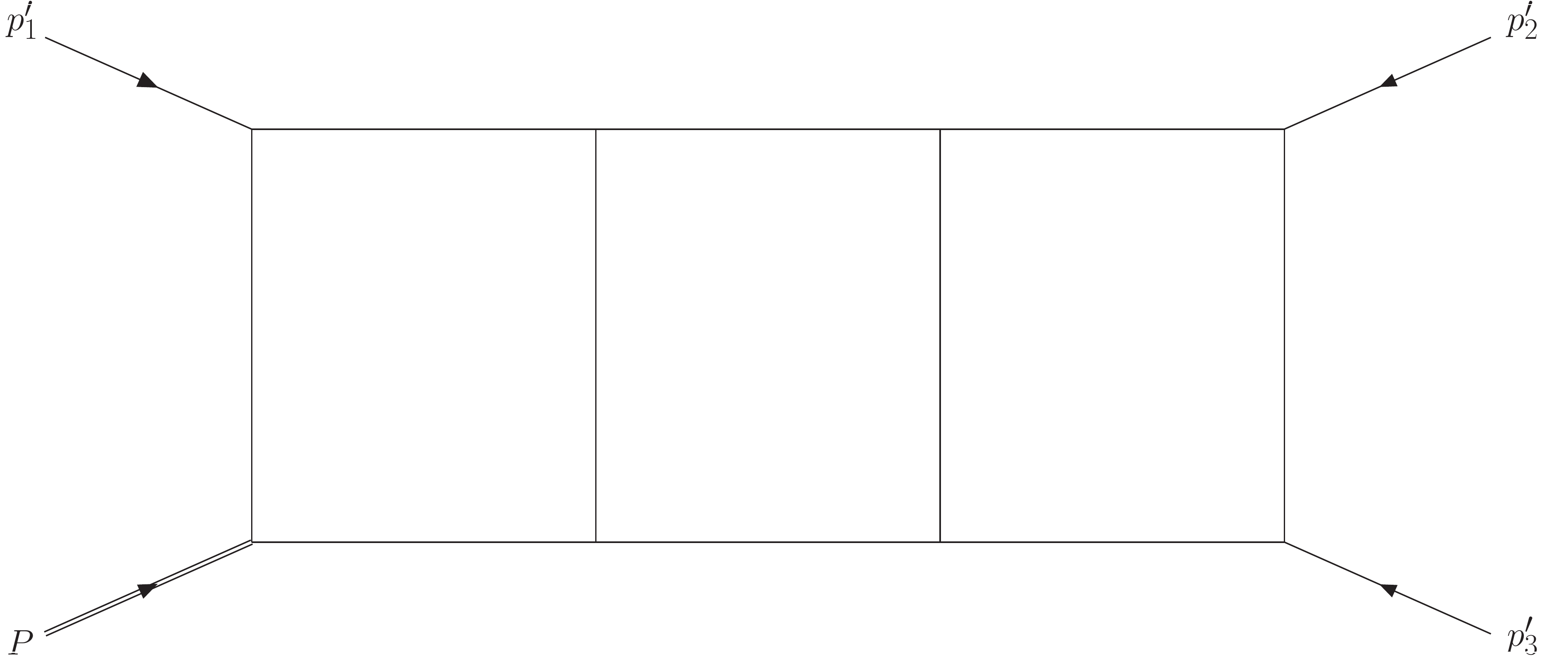}
\caption{The ladderbox (F1) family using the notation followed in \cite{Chicherin:2020umh}. The double line represents the massive particle and all external momenta are taken to be incoming.}
\label{figure2}
\end{figure}

For the standard DE method we are going to use the notation used in \cite{ Chicherin:2020umh} (see figure \ref{figure2}). In this article, the Lorentz invariants are defined as follows  
\begin{equation}
\label{PapathaKin}
z_1=\frac{2 p'_1 \cdot p'_2}{P^2}, \, \, \, \, z_2=\frac{2 p'_2 \cdot p'_3}{P^2}\, \, \, \, \text{and} \, \, \, \, z_3=1-z_1-z_2 .
\end{equation}
The alphabet of this family (using the standard notation \cite{Duhr:2011zq, Duhr:2012fh, Duhr:2014woa}), as it is derived by applying the standard DE method, is
\begin{equation}
\label{PapathaAlpha}
\{ z_1,z_2,z_3,1-z_1,1-z_2,1-z_3 \} \to \{ z_1,z_2,1-z_1-z_2,1-z_1,1-z_2,z_1+z_2 \}
\end{equation}
and the pure basis of master integrals $\textbf{G}$ satisfies the differential equation
\begin{equation}
\label{DE}
\begin{split}
\mathrm{d} \textbf{G}&=\epsilon \bigg(\textbf{A}_{z_1} \mathrm{d} \log(z_1)+ \textbf{A}_{z_2} \mathrm{d} \log(z_2)+ \textbf{A}_{1-z_1-z_2} \mathrm{d} \log(1-z_1-z_2) \\
&+\textbf{A}_{1-z_1} \mathrm{d} \log(1-z_1)+\textbf{A}_{1-z_2} \mathrm{d} \log(1-z_2)+\textbf{A}_{z_1+z_2} \mathrm{d} \log(z_1+z_2)\bigg)  \textbf{G}
\end{split}
\end{equation}
The $C_2$ adjacency conditions \cite{Chicherin:2020umh} imply that the letters $\{1-z_i,1-z_j\}$ for $i,j=1,2,3$ and $i \neq j$ never appear next to each other in a symbol, or equivalently (in terms of the matrices of the DE), that the matrices of the DE satisfy the following conditions
\begin{equation}
\label{AdjaceRela}
\textbf{A}_{1-z_1}\textbf{A}_{1-z_2}=0, \, \, \, \, \textbf{A}_{1-z_1}\textbf{A}_{z_1+z_2}=0 \, \, \, \, \text{and} \, \, \, \, \textbf{A}_{1-z_2}\textbf{A}_{z_1+z_2}=0.
\end{equation}

As we have already discussed, in our treatment of the three-loop planar families with one off-shell leg, we use the notation as is depicted in figure \ref{fig:topgraphs}, while the kinematic invariants are defined as in \eqref{eq:invmap} with the extra introduction of the ratio $y=s_{23}/s_{12}$ for convenience. The pure basis of master integrals satisfies the differential equation \eqref{eq:cansde}, which in more detail can be written as
\begin{equation}
\label{SDE}
\frac{\mathrm{d}}{\mathrm{d}x}\textbf{G}=\epsilon \left( \frac{\textbf{M}_0}{x}+\frac{\textbf{M}_1}{x-1}+\frac{\textbf{M}_{-1/y}}{x+1/y}+\frac{\textbf{M}_{1/(1+y)}}{x-1/(1+y)} \right)\textbf{G}.
\end{equation}

From figures \ref{fig:topgraphs} and \ref{figure2} (using also \eqref{eq:xparam}) we observe that the momenta in the two different notations are related via the transformation
\begin{equation}
\label{MomeTran}
P\to p_1+p_2-x p_1, \, \, \, \, p'_1\to x p_1, \, \, \, \, p'_2\to p_4, \, \, \, \, \text{and} \, \, \, \, p'_3\to p_3.
\end{equation}
Thus using \eqref{MomeTran} in \eqref{PapathaKin} we find that the transformation that takes us from the $\{z_1,z_2\}$ invariants to the $\{x,y\}$ ones, is
\begin{equation}
\label{ToSDE}
z_1=\frac{x y}{1-x} \, \, \, \, \text{and} \, \, \, \, z_2=\frac{1}{1-x}
\end{equation}
while the reverse transformation reads
\begin{equation}
\label{FromSDE}
x=\frac{z_2-1}{z_2} \, \, \, \, \text{and} \, \, \, \, y=\frac{z_1}{z_2-1}.
\end{equation}
In addittion, using \eqref{ToSDE} the alphabet in \eqref{PapathaAlpha} can be written in terms of $\{x,y\}$ as (keeping the same ordering for the letters)
\begin{equation}
\label{lettersxy}
\left\{\frac{x y}{1-x}, \frac{1}{1-x}, \frac{-x(1+y)}{1-x},-\left(\frac{1+y}{1-x}\right)\left(x-\frac{1}{1+y}\right),\frac{-x}{1-x},\left(\frac{y}{1-x}\right)\left(x+\frac{1}{y}\right) \right\}.
\end{equation}

Having established the connection between the standard DE method and the SDE approach, via the information stated above, we can see now how the adjacency conditions are interpreted within the latter. More specifically, expressing the letters in \eqref{DE} via the relation \eqref{lettersxy} and differentiating with respect to $x$ we obtain
\begin{equation}
\label{SDElike}
\begin{split}
    \frac{\mathrm{d}}{\mathrm{d}x}\textbf{G}&=\epsilon \left(\frac{\textbf{A}_{1-z_2}+\textbf{A}_{z_1}+ \textbf{A}_{1-z_1-z_2}}{x} +\frac{\textbf{A}_{z_1+z_2}}{x+1/y}+\frac{\textbf{A}_{1-z_1}}{x-1/(1+y)} \right. \\
    &\left. -\frac{\textbf{A}_{z_1}+\textbf{A}_{z_2}+\textbf{A}_{1-z_1-z_2} +\textbf{A}_{1-z_1} +\textbf{A}_{1-z_2} +\textbf{A}_{z_1+z_2}}{x-1}  \right)\textbf{G}
\end{split}
\end{equation}
Comparing equations \eqref{SDE} and \eqref{SDElike}, we find that following relations arise among the residue matrices of the standard DE method and the SDE approach
\begin{equation}
\label{relations}
\begin{split}
&\textbf{A}_{1-z_1}=\textbf{M}_{1/(1+y)} \\
&\textbf{A}_{z_1+z_2}=\textbf{M}_{-1/y} \\
&\textbf{A}_{z_2}=-\textbf{M}_0-\textbf{M}_1-\textbf{M}_{1/(1+y)}-\textbf{M}_{1/(1+y)} \\
&\textbf{A}_{1-z_2}=\textbf{M}_{0}-\textbf{A}_{z_1}-\textbf{A}_{1-z_1-z_2}
\end{split}
\end{equation}
From \eqref{relations} we see that the matrices $\{\textbf{A}_{1-z_1},\textbf{A}_{z_1+z_2\}}$ are completely determined by the residue matrices of SDE, while $\textbf{A}_{1-z_2}$ remains unknown.  Thus the only adjacency relation that can be passed from the DE method to the SDE residue matrices is the
\begin{equation}
     \textbf{A}_{1-z_1}\textbf{A}_{z_1+z_2}=0 \Rightarrow \textbf{M}_{1/(1+y)} \textbf{M}_{-1/y} =0
\end{equation}
Having the residue matrices $\{ \textbf{M}_{1/(1+y)},\textbf{M}_{-1/y}\}$ for the families F1, F2 and F3 we checked and we found out that indeed the above relation holds true for all of them\footnote{For the F1 family this was obvious, as the adjacency conditions have been already studied for this family.}. Moreover, assuming that (as in the SDE approach) the alphabet of the two tennis-court families is the same with the one of the ladder-box family, this result can be viewed as a first indication of the validity of the adjacency conditions for the tennis-court families.


\begin{thebibliography}{110}
\justifying
\bibitem{Gehrmann:2021qex}
T.~Gehrmann and B.~Malaescu,
``Precision QCD Physics at the LHC,''
\href{https://arxiv.org/abs/2111.02319}{[arXiv:2111.02319 [hep-ph]]}.

\bibitem{Amoroso:2020lgh}
S.~Amoroso, P.~Azzurri, J.~Bendavid, E.~Bothmann, D.~Britzger, H.~Brooks, A.~Buckley, M.~Calvetti, X.~Chen and M.~Chiesa, \textit{et al.}
``Les Houches 2019: Physics at TeV Colliders: Standard Model Working Group Report,''
\href{https://arxiv.org/abs/2003.01700}{[arXiv:2003.01700 [hep-ph]]}.

\bibitem{Heinrich:2020ybq}
G.~Heinrich,
``Collider Physics at the Precision Frontier,''
Phys. Rept. \textbf{922} (2021), 1-69
doi:10.1016/j.physrep.2021.03.006
\href{https://arxiv.org/abs/2009.00516}{[arXiv:2009.00516 [hep-ph]]}.

\bibitem{Tancredi:2021oiq}
L.~Tancredi,
``Calculational Techniques in Particle Theory,''
\href{https://arxiv.org/abs/2111.00205}{[arXiv:2111.00205 [hep-ph]]}.


\bibitem{Henn:2013tua}
J.~M.~Henn, A.~V.~Smirnov and V.~A.~Smirnov,
``Analytic results for planar three-loop four-point integrals from a Knizhnik-Zamolodchikov equation,''
JHEP \textbf{07} (2013), 128
doi:10.1007/JHEP07(2013)128
\href{https://arxiv.org/abs/1306.2799}{[arXiv:1306.2799 [hep-th]]}.

\bibitem{Henn:2020lye}
J.~Henn, B.~Mistlberger, V.~A.~Smirnov and P.~Wasser,
``Constructing d-log integrands and computing master integrals for three-loop four-particle scattering,''
JHEP \textbf{04} (2020), 167
doi:10.1007/JHEP04(2020)167
\href{https://arxiv.org/abs/2002.09492}{[arXiv:2002.09492 [hep-ph]]}.

\bibitem{Caola:2020dfu}
F.~Caola, A.~Von Manteuffel and L.~Tancredi,
``Diphoton Amplitudes in Three-Loop Quantum Chromodynamics,''
Phys. Rev. Lett. \textbf{126} (2021) no.11, 112004
doi:10.1103/PhysRevLett.126.112004
\href{https://arxiv.org/abs/2011.13946}{[arXiv:2011.13946 [hep-ph]]}.

\bibitem{Caola:2021rqz}
F.~Caola, A.~Chakraborty, G.~Gambuti, A.~von Manteuffel and L.~Tancredi,
``Three-loop helicity amplitudes for four-quark scattering in massless QCD,''
JHEP \textbf{10} (2021), 206
doi:10.1007/JHEP10(2021)206
\href{https://arxiv.org/abs/2108.00055}{[arXiv:2108.00055 [hep-ph]]}.

\bibitem{Bargiela:2021wuy}
P.~Bargiela, F.~Caola, A.~von Manteuffel and L.~Tancredi,
``Three-loop helicity amplitudes for diphoton production in gluon fusion,''
\href{https://arxiv.org/abs/2111.13595}{[arXiv:2111.13595 [hep-ph]]}.

\bibitem{Caola:2021izf}
F.~Caola, A.~Chakraborty, G.~Gambuti, A.~von Manteuffel and L.~Tancredi,
``Three-loop gluon scattering in QCD and the gluon Regge trajectory,''
\href{https://arxiv.org/abs/2112.11097}{[arXiv:2112.11097 [hep-ph]]}.

\bibitem{de1}
A.~V.~Kotikov,
``Differential equations method: New technique for massive Feynman diagrams calculation,''
Phys. Lett. B \textbf{254} (1991), 158-164.

\bibitem{de2}
A.~V.~Kotikov,
``Differential equations method: The Calculation of vertex type Feynman diagrams,''
Phys. Lett. B \textbf{259} (1991), 314-322.

\bibitem{de3}
A.~V.~Kotikov,
``Differential equation method: The Calculation of N point Feynman diagrams,''
Phys. Lett. B \textbf{267} (1991), 123-127 [Errattum: Phys. Lett. B \textbf{295} (1992), 409].

\bibitem{de4}
T.~Gehrmann and E.~Remiddi,
``Differential equations for two loop four point functions,''
Nucl. Phys. B \textbf{580} (2000), 485-518
\href{https://arxiv.org/abs/hep-ph/9912329}{[arXiv:hep-ph/9912329 [hep-ph]]}.

\bibitem{Gehrmann:2000zt}
T.~Gehrmann and E.~Remiddi,
``Two loop master integrals for gamma* ---\ensuremath{>} 3 jets: The Planar topologies,''
Nucl. Phys. B \textbf{601} (2001), 248-286
doi:10.1016/S0550-3213(01)00057-8
\href{https://arxiv.org/abs/hep-ph/0008287}{[arXiv:hep-ph/0008287 [hep-ph]]}.

\bibitem{Gehrmann:2001ck}
T.~Gehrmann and E.~Remiddi,
``Two loop master integrals for gamma* --\ensuremath{>} 3 jets: The Nonplanar topologies,''
Nucl. Phys. B \textbf{601} (2001), 287-317
doi:10.1016/S0550-3213(01)00074-8
\href{https://arxiv.org/abs/hep-ph/0101124}{[arXiv:hep-ph/0101124 [hep-ph]]}.

\bibitem{Chetyrkin:1981qh}
K.~G.~Chetyrkin and F.~V.~Tkachov,
``Integration by Parts: The Algorithm to Calculate beta Functions in 4 Loops,''
Nucl. Phys. B \textbf{192} (1981), 159-204
doi:10.1016/0550-3213(81)90199-1.

\bibitem{Laporta:2000dsw}
S.~Laporta,
``High precision calculation of multiloop Feynman integrals by difference equations,''
Int. J. Mod. Phys. A \textbf{15} (2000), 5087-5159
doi:10.1142/S0217751X00002159
\href{https://arxiv.org/abs/hep-ph/0102033}{[arXiv:hep-ph/0102033 [hep-ph]]}.

\bibitem{DiVita:2014pza}
S.~Di Vita, P.~Mastrolia, U.~Schubert and V.~Yundin,
``Three-loop master integrals for ladder-box diagrams with one massive leg,''
JHEP \textbf{09} (2014), 148
doi:10.1007/JHEP09(2014)148
\href{https://arxiv.org/abs/1408.3107}{[arXiv:1408.3107 [hep-ph]]}.

\bibitem{Henn:2013pwa}
J.~M.~Henn,
``Multiloop integrals in dimensional regularization made simple,''
Phys. Rev. Lett. \textbf{110} (2013), 251601
doi:10.1103/PhysRevLett.110.251601
\href{https://arxiv.org/abs/1304.1806}{[arXiv:1304.1806 [hep-th]]}.

\bibitem{Arkani-Hamed:2010pyv}
N.~Arkani-Hamed, J.~L.~Bourjaily, F.~Cachazo and J.~Trnka,
``Local Integrals for Planar Scattering Amplitudes,''
JHEP \textbf{06} (2012), 125
doi:10.1007/JHEP06(2012)125
\href{https://arxiv.org/abs/1012.6032}{[arXiv:1012.6032 [hep-th]]}.

\bibitem{Argeri:2014qva}
M.~Argeri, S.~Di Vita, P.~Mastrolia, E.~Mirabella, J.~Schlenk, U.~Schubert and L.~Tancredi,
``Magnus and Dyson Series for Master Integrals,''
JHEP \textbf{03} (2014), 082
doi:10.1007/JHEP03(2014)082
\href{https://arxiv.org/abs/1401.2979}{[arXiv:1401.2979 [hep-ph]]}.

\bibitem{Canko:2020gqp}
D.~D.~Canko and N.~Syrrakos,
``Resummation methods for Master Integrals,''
JHEP \textbf{02} (2021), 080
doi:10.1007/JHEP02(2021)080
\href{https://arxiv.org/abs/2010.06947}{[arXiv:2010.06947 [hep-ph]]}.

\bibitem{Papadopoulos:2014lla}
C.~G.~Papadopoulos,
``Simplified differential equations approach for Master Integrals,''
JHEP \textbf{07} (2014), 088
doi:10.1007/JHEP07(2014)088 
\href{https://arxiv.org/abs/1401.6057}{[arXiv:1401.6057 [hep-ph]]}.

\bibitem{Papadopoulos:2015jft}
C.~G.~Papadopoulos, D.~Tommasini and C.~Wever,
``The Pentabox Master Integrals with the Simplified Differential Equations approach,''
JHEP \textbf{04} (2016), 078
doi:10.1007/JHEP04(2016)078
\href{https://arxiv.org/abs/1511.09404}{[arXiv:1511.09404 [hep-ph]]}.

\bibitem{Smirnov:2019qkx}
A.~V.~Smirnov and F.~S.~Chuharev,
Comput. Phys. Commun. \textbf{247 } (2020), 106877
doi:10.1016/j.cpc.2019.106877
\href{https://arxiv.org/abs/1901.07808}{[arXiv:1901.07808 [hep-ph]]}.

\bibitem{Klappert:2020nbg}
J.~Klappert, F.~Lange, P.~Maierh\"ofer and J.~Usovitsch,
``Integral reduction with Kira 2.0 and finite field methods,''
Comput. Phys. Commun. \textbf{266} (2021), 108024
doi:10.1016/j.cpc.2021.108024
\href{https://arxiv.org/abs/2008.06494}{[arXiv:2008.06494 [hep-ph]]}.

\bibitem{Goncharov:1998kja}
A.~B.~Goncharov,
``Multiple polylogarithms, cyclotomy and modular complexes,''
Math. Res. Lett. \textbf{5} (1998), 497-516
doi:10.4310/MRL.1998.v5.n4.a7
\href{https://arxiv.org/abs/1105.2076}{[arXiv:1105.2076 [math.AG]]}.

\bibitem{Duhr:2011zq}
C.~Duhr, H.~Gangl and J.~R.~Rhodes,
``From polygons and symbols to polylogarithmic functions,''
JHEP \textbf{10} (2012), 075
doi:10.1007/JHEP10(2012)075
\href{https://arxiv.org/abs/1110.0458}{[arXiv:1110.0458 [math-ph]]}.

\bibitem{Duhr:2012fh}
C.~Duhr,
``Hopf algebras, coproducts and symbols: an application to Higgs boson amplitudes,''
JHEP \textbf{08} (2012), 043
doi:10.1007/JHEP08(2012)043
\href{https://arxiv.org/abs/1203.0454}{[arXiv:1203.0454 [hep-ph]]}.

\bibitem{Duhr:2014woa}
C.~Duhr,
``Mathematical aspects of scattering amplitudes,''
doi:10.1142/9789814678766\_0010
\href{https://arxiv.org/abs/1411.7538}{[arXiv:1411.7538 [hep-ph]]}.



\bibitem{Dixon:2020bbt}
L.~J.~Dixon, A.~J.~McLeod and M.~Wilhelm,
``A Three-Point Form Factor Through Five Loops,''
JHEP \textbf{04} (2021), 147
doi:10.1007/JHEP04(2021)147
\href{https://arxiv.org/abs/2012.12286}{[arXiv:2012.12286 [hep-th]]}.

\bibitem{Chicherin:2020umh}
D.~Chicherin, J.~M.~Henn and G.~Papathanasiou,
``Cluster algebras for Feynman integrals,''
Phys. Rev. Lett. \textbf{126} (2021) no.9, 091603
doi:10.1103/PhysRevLett.126.091603
\href{https://arxiv.org/abs/2012.12285}{[arXiv:2012.12285 [hep-th]]}.


\bibitem{Bogner:2010kv}
C.~Bogner and S.~Weinzierl,
``Feynman graph polynomials,''
Int. J. Mod. Phys. A \textbf{25} (2010), 2585-2618
doi:10.1142/S0217751X10049438
\href{https://arxiv.org/abs/1002.3458}{[arXiv:1002.3458 [hep-ph]]}.

\bibitem{Panzer:2014caa}
E.~Panzer,
``Algorithms for the symbolic integration of hyperlogarithms with applications to Feynman integrals,''
Comput. Phys. Commun. \textbf{188} (2015), 148-166
doi:10.1016/j.cpc.2014.10.019
\href{https://arxiv.org/abs/1403.3385}{[arXiv:1403.3385 [hep-th]]}.

\bibitem{Duhr:2019tlz}
C.~Duhr and F.~Dulat,
``PolyLogTools \textemdash{} polylogs for the masses,''
JHEP \textbf{08} (2019), 135
doi:10.1007/JHEP08(2019)135
\href{https://arxiv.org/abs/1904.07279}{[arXiv:1904.07279 [hep-th]]}.

\bibitem{Meyer:2017joq}
C.~Meyer,
``Algorithmic transformation of multi-loop master integrals to a canonical basis with CANONICA,''
Comput. Phys. Commun. \textbf{222} (2018), 295-312
doi:10.1016/j.cpc.2017.09.014
\href{https://arxiv.org/abs/1705.06252}{[arXiv:1705.06252 [hep-ph]]}.

\bibitem{Prausa:2017ltv}
M.~Prausa,
``epsilon: A tool to find a canonical basis of master integrals,''
Comput. Phys. Commun. \textbf{219} (2017), 361-376
doi:10.1016/j.cpc.2017.05.026
\href{https://arxiv.org/abs/1701.00725}{[arXiv:1701.00725 [hep-ph]]}.

\bibitem{Gituliar:2017vzm}
O.~Gituliar and V.~Magerya,
``Fuchsia: a tool for reducing differential equations for Feynman master integrals to epsilon form,''
Comput. Phys. Commun. \textbf{219} (2017), 329-338
doi:10.1016/j.cpc.2017.05.004
\href{https://arxiv.org/abs/1701.04269}{[arXiv:1701.04269 [hep-ph]]}.

\bibitem{Dlapa:2020cwj}
C.~Dlapa, J.~Henn and K.~Yan,
``Deriving canonical differential equations for Feynman integrals from a single uniform weight integral,''
JHEP \textbf{05} (2020), 025
doi:10.1007/JHEP05(2020)025
\href{https://arxiv.org/abs/2002.02340}{[arXiv:2002.02340 [hep-ph]]}.

\bibitem{Wasser:2018qvj}
P.~Wasser,
``Analytic properties of Feynman integrals for scattering amplitudes,''

\bibitem{Henn:2014lfa}
J.~M.~Henn, K.~Melnikov and V.~A.~Smirnov,
``Two-loop planar master integrals for the production of off-shell vector bosons in hadron collisions,''
JHEP \textbf{05} (2014), 090
doi:10.1007/JHEP05(2014)090
\href{https://arxiv.org/abs/1402.7078}{[arXiv:1402.7078 [hep-ph]]}.

\bibitem{Dlapa:2021qsl}
C.~Dlapa, X.~Li and Y.~Zhang,
``Leading singularities in Baikov representation and Feynman integrals with uniform transcendental weight,''
doi:10.1007/JHEP07(2021)227
\href{https://arxiv.org/abs/2103.04638}{[arXiv:2103.04638 [hep-th]]}.

\bibitem{Canko:2021hvh}
D.~D.~Canko, F.~Gasparotto, L.~Mattiazzi, C.~G.~Papadopoulos and N.~Syrrakos,
``$N^3LO$ calculations for $2 \to 2$ processes using Simplified Differential Equations,''
\href{https://arxiv.org/abs/2110.08110}{[arXiv:2110.08110 [hep-ph]]}.

\bibitem{Canko:2020ylt}
D.~D.~Canko, C.~G.~Papadopoulos and N.~Syrrakos,
``Analytic representation of all planar two-loop five-point Master Integrals with one off-shell leg,''
JHEP \textbf{01} (2021), 199
doi:10.1007/JHEP01(2021)199
\href{https://arxiv.org/abs/2009.13917}{[arXiv:2009.13917 [hep-ph]]}.

\bibitem{Dulat:2014mda}
F.~Dulat and B.~Mistlberger,
``Real-Virtual-Virtual contributions to the inclusive Higgs cross section at N3LO,''
\href{https://arxiv.org/abs/1411.3586}{[arXiv:1411.3586 [hep-ph]]}.

\bibitem{Mastrolia:2017pfy}
P.~Mastrolia, M.~Passera, A.~Primo and U.~Schubert,
``Master integrals for the NNLO virtual corrections to $\mu e$ scattering in QED: the planar graphs,''
JHEP \textbf{11} (2017), 198
doi:10.1007/JHEP11(2017)198
\href{https://arxiv.org/abs/1709.07435}{[arXiv:1709.07435 [hep-ph]]}.
\bibitem{Jantzen:2012mw}
B.~Jantzen, A.~V.~Smirnov and V.~A.~Smirnov,
``Expansion by regions: revealing potential and Glauber regions automatically,''
Eur. Phys. J. C \textbf{72} (2012), 2139
doi:10.1140/epjc/s10052-012-2139-2
\href{https://arxiv.org/abs/1206.0546}{[arXiv:1206.0546 [hep-ph]]}.

\bibitem{Smirnov:2015mct}
A.~V.~Smirnov,
``FIESTA4: Optimized Feynman integral calculations with GPU support,''
Comput. Phys. Commun. \textbf{204} (2016), 189-199
doi:10.1016/j.cpc.2016.03.013
\href{https://arxiv.org/abs/1511.03614}{[arXiv:1511.03614 [hep-ph]]}.

\bibitem{Syrrakos:2020kba}
N.~Syrrakos,
``Pentagon integrals to arbitrary order in the dimensional regulator,''
JHEP \textbf{06} (2021), 037
doi:10.1007/JHEP06(2021)037
\href{https://arxiv.org/abs/2012.10635}{[arXiv:2012.10635 [hep-ph]]}.

\bibitem{Syrrakos:2021nij}
N.~Syrrakos,
``One-loop Feynman integrals for 2 \textrightarrow{} 3 scattering involving many scales including internal masses,''
JHEP \textbf{10} (2021), 041
doi:10.1007/JHEP10(2021)041
\href{https://arxiv.org/abs/2107.02106}{[arXiv:2107.02106 [hep-ph]]}.

\bibitem{pslq}
H.R.P. Ferguson and D.H. Bailey, \textit{A polynomial time, numerically stable integer relation algorithm}, RNR technical report RNR-91-032, (1992).

\bibitem{Vollinga:2004sn}
J.~Vollinga and S.~Weinzierl,
``Numerical evaluation of multiple polylogarithms,''
Comput. Phys. Commun. \textbf{167} (2005), 177
doi:10.1016/j.cpc.2004.12.009
\href{https://arxiv.org/abs/hep-ph/0410259}{[arXiv:hep-ph/0410259 [hep-ph]]}.

\bibitem{Borowka:2017idc}
S.~Borowka, G.~Heinrich, S.~Jahn, S.~P.~Jones, M.~Kerner, J.~Schlenk and T.~Zirke,
``pySecDec: a toolbox for the numerical evaluation of multi-scale integrals,''
Comput. Phys. Commun. \textbf{222} (2018), 313-326
doi:10.1016/j.cpc.2017.09.015
\href{https://arxiv.org/abs/1703.09692}{[arXiv:1703.09692 [hep-ph]]}.

\bibitem{gitfiles}
\href{https://github.com/nsyrrakos/Planar3loops1mass.git}{https://github.com/nsyrrakos/Planar3loops1mass.git}.


\bibitem{Binosi:2008ig}
D.~Binosi, J.~Collins, C.~Kaufhold and L.~Theussl,
``JaxoDraw: A Graphical user interface for drawing Feynman diagrams. Version 2.0 release notes,''
Comput. Phys. Commun. \textbf{180} (2009), 1709-1715
doi:10.1016/j.cpc.2009.02.020
\href{https://arxiv.org/abs/0811.4113}{[arXiv:0811.4113 [hep-ph]]}.


\end{thebibliography}
\end{document}